\documentclass{PoS}

\title{Comparison between LQCD and PNJL model at finite chemical potentials}

\ShortTitle{Comparison between LQCD and PNJL at finite chemical potentials}

\author{\speaker{Yuji Sakai}$^{1}$,  
{Takahiro Sasaki$^{1}$, Hiroaki Kouno$^{2}$, Masanobu Yahiro$^{1}$}\\
        $^{1}$Department of Physics, Kyushu University, 
	         Fukuoka 812-8581, Japan\\
	$^{2}$Department of Physics, Saga University, 
	          Saga 840-8502, Japan\\
        $^*$E-mail: \email{sakai@phys.kyushu-u.ac.jp}}

\abstract{Lattice QCD (LQCD) has the sign problem at real quark chemical 
potential. There are some regions with no sign problem; one is the imaginary 
quark chemical potential region and the others are the real and imaginary 
isospin chemical potential regions. 
We show that the Polyakov-loop extended Nambu--Jona-Lasinio (PNJL) model can 
reproduce LQCD data in the regions. We also determine the model parameters 
from the data and predict the QCD phase diagram in the real quark 
chemical potential region.}

\FullConference{The XXVIII International Symposium on Lattice Field Theory, Lattice2010\\
		June 14-19, 2010\\
		Villasimius, Italy}

\begin{document}

\section{Introduction}

The QCD phase diagram is essential for understanding not only natural 
phenomena such as compact stars and the early universe but also laboratory 
experiments such as relativistic heavy-ion collisions.
Quantitative calculations of the phase diagram from the first-principle 
lattice QCD (LQCD) have the sign problem at real quark chemical potential 
($\mu_{\rm q}$). 
Though several approaches have been proposed to circumvent the difficulty, 
these are still far from perfection.

As an approach complementary to the first-principle LQCD, we can consider 
effective models such as the Nambu--Jona-Lasinio (NJL) model and the 
Polyakov-loop extended Nambu--Jona-Lasinio (PNJL) model. 
The NJL model describes the chiral symmetry breaking, but not the confinement 
mechanism. The PNJL model is constructed so as to treat both the mechanisms.
In the NJL-type models, the input parameters are determined at $\mu_{\rm q}=0$.
It is then highly nontrivial whether the models predict the dynamics of QCD at 
finite $\mu_{\rm q}$ properly. This should be tested from QCD. 
Fortunately, this is possible in some regions without sign problem, 
such as imaginary $\mu_{\rm q}$, real and imaginary isospin chemical potential 
($\mu_{\rm I}$).

In this paper, we consider two-flavor QCD and show the reliability of the PNJL 
model by comparing the model result with LQCD data in their regions. 

\section{Imaginary Quark Chemical Potential}

Roberge and Weiss~\cite{RW} found that the thermodynamic potential, 
$\Omega_{\rm QCD}(\theta_{\rm q})$, of QCD at imaginary chemical potential 
$\mu_{\rm q}=i\theta_{\rm q}T$ has a periodicity 
$\Omega_{\rm QCD}(\theta_{\rm q}) = \Omega_{\rm QCD}(\theta_{\rm q} + 2\pi k/3)$, showing that $\Omega_{\rm QCD}(\theta_{\rm q} + 2\pi k/3)$ is transformed 
into $\Omega_{\rm QCD}(\theta_{\rm q})$ by the $\mathbb{Z}_3$ transformation 
with integer $k$. 
This means that QCD is invariant under a combination of the $\mathbb{Z}_3$ 
transformation and a parameter transformation 
$\theta_{\rm q}\rightarrow\theta_{\rm q}+2\pi k/3$. 
We call this combination the extended $\mathbb{Z}_3$ transformation. 
Thus, $\Omega_{\rm QCD}(\theta_{\rm q})$ has the extended $\mathbb{Z}_3$ 
symmetry, and hence quantities invariant under the extended $\mathbb{Z}_3$ 
transformation have the RW periodicity~\cite{Sakai-1}.

We reveal that the PNJL model has the RW periodicity~\cite{Sakai-1}.
The two-flavor PNJL Lagrangian~\cite{Fukushima} in Euclidean spacetime is
%%%%%%%%%%%%%%%%%%%%%%%%%%%%%%%%%%%%%%%%%%%%%%%%%%%%%%%%%%%%%%%%%%
\begin{eqnarray}
{\cal L}=\bar{q}(i\gamma_\nu D_\nu-\gamma_4\mu_{\rm q}+m_0)q
-G_{\rm s}[(\bar{q}q)^2+
(\bar{q}i\gamma_5\vec{\tau}q)^2]+U_{\Phi}(\Phi[A],\Phi^*[A],T), 
\label{Eq2-1}
\end{eqnarray}
%%%%%%%%%%%%%%%%%%%%%%%%%%%%%%%%%%%%%%%%%%%%%%%%%%%%%%%%%%%%%%%%%%
where $q$ denotes the two-flavor quark field, $m_0$ does the current quark mass, and $D_{\nu}=\partial_{\nu}-iA_{\nu}\delta_{\nu 0}$ with the gauge field 
$A_{\nu}$. 
In the chiral limit ($m_0=0$), the Lagrangian density has the exact 
$SU(2)_{\rm R}\times SU(2)_{\rm L}\times U(1)_{\rm v}\times SU(3)_{\rm c}$ 
symmetry. 
The Polyakov potential $U_{\Phi}$~\cite{Rossner} is a function of the Polyakov 
loop $\Phi=\frac{1}{3}{\rm tr_c}~L$ with $L=e^{iA_4/T}$ and its Hermitian 
conjugate $\Phi^*$. 
The PNJL thermodynamic potential $\Omega$ in the mean field approximation 
(MFA) is 
%%%%%%%%%%%%%%%%%%%%%%%%%%%%%%%%%%%%%%%%%%%%%%%%%%%%%%%%%%%%%%%%%%
\begin{eqnarray}
\Omega=-4\int\frac{d^3{\bf p}}{(2\pi)^3}
\Bigl[3\epsilon({\bf p})+T\sum_{\lambda=\pm 1}
{\rm ln~det_c}(1+L^{\lambda}e^{-\epsilon({\bf p})/T+i\lambda\theta_{\rm q}})
\Bigr]+G_{\rm s}\sigma^2+U_{\Phi},
\label{Eq2-2}
\end{eqnarray}
%%%%%%%%%%%%%%%%%%%%%%%%%%%%%%%%%%%%%%%%%%%%%%%%%%%%%%%%%%%%%%%%%%
where $\epsilon=\sqrt{{\bf p}^2+M^2}$, $M=m_0-2G_{\rm s}\sigma$, 
and $\sigma=\langle\bar{q}q\rangle$. 
The thermodynamic potential $\Omega$ is invariant 
under the extended $\mathbb{Z}_3$ transformation, 
%%%%%%%%%%%%%%%%%%%%%%%%%%%%%%%%%%%%%%%%%%%%%%%%%%%%%%%%%%%%%%%%%%
\begin{eqnarray}
L\rightarrow e^{-i2\pi k/3}L,~~
\theta_{\rm q}\rightarrow\theta_{\rm q}+2\pi k/3. 
\label{Eq2-3}
\end{eqnarray}
%%%%%%%%%%%%%%%%%%%%%%%%%%%%%%%%%%%%%%%%%%%%%%%%%%%%%%%%%%%%%%%%%%
Therefore, $\Omega$ has the RW periodicity.

%%%%%%%%%%%%%%%%%%%%%%%%%%%%%%%%%%%%%%%%%%%%%%%%%%%%%%%%%%%%
\begin{figure}[htbp]
\begin{center}
 \includegraphics[width=0.4\textwidth,angle=-90]{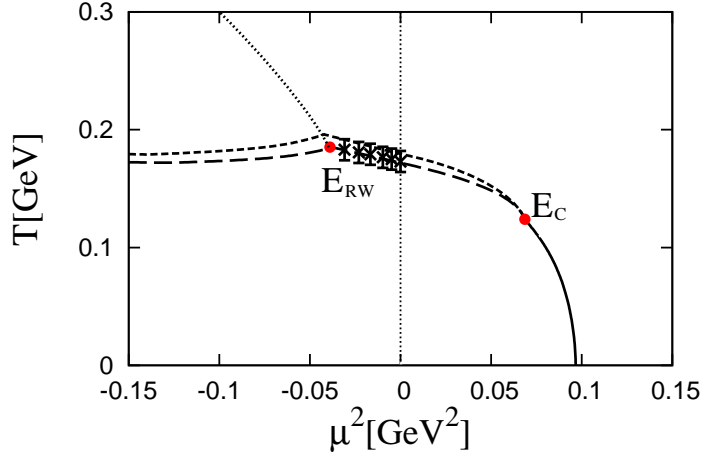}
\end{center}
\caption{
Phase diagram in the real and imaginary $\mu_{\rm q}$ regions by 
the PNJL model with the parameter set~\cite{Sakai-1} that reproduces 
the LQCD data at imaginary $\mu_{\rm q}$. 
The points ${\rm E_{RW}}$ and ${\rm E_{C}}$ are the endpoints 
of the RW transition and the first-order chiral transition 
respectively. 
The solid line denotes the first-order chiral transition, 
the dashed (dotted) line does the crossover deconfinement (chiral) 
transition, and the dot-dashed line does the RW transition. 
Lattice data ($\times$) are taken from~\cite{LQCD-1}.}
\label{Fig1}
\end{figure}
%%%%%%%%%%%%%%%%%%%%%%%%%%%%%%%%%%%%%%%%%%%%%%%%%%%%%%%%%%%%

At the present stage, the PNJL model is only a realistic effective model that 
possesses both the extended $\mathbb{Z}_3$ symmetry and the chiral symmetry
~\cite{Sakai-1}. 
This property guarantees that the phase diagram evaluated by the PNJL model 
has the RW periodicity in the imaginary $\mu_{\rm q}$ region, and therefore 
makes it possible to compare the PNJL result with LQCD data 
quantitatively in the imaginary $\mu_{\rm q}$ region. 
Actually, the PNJL model succeeds in reproducing the LQCD data~\cite{LQCD-1} 
by introducing the vector-type four-quark interaction and the scalar-type 
eight-quark interaction~\cite{Sakai-1}. 
The QCD phase diagram in the real $\mu_{\rm q}$ region is 
predicted by the PNJL model with the parameter set~\cite{Sakai-1} 
that reproduces the LQCD data at imaginary $\mu_{\rm q}$, as shown in 
Fig.~\ref{Fig1}.

%%%%%%%%%%%%%%%%%%%%%%%%%%%%%%%%%%%%%%%%%%%%%%%%%%%%%%%%%%%%
\begin{figure}[htbp]
\begin{center}
 \includegraphics[width=0.4\textwidth]{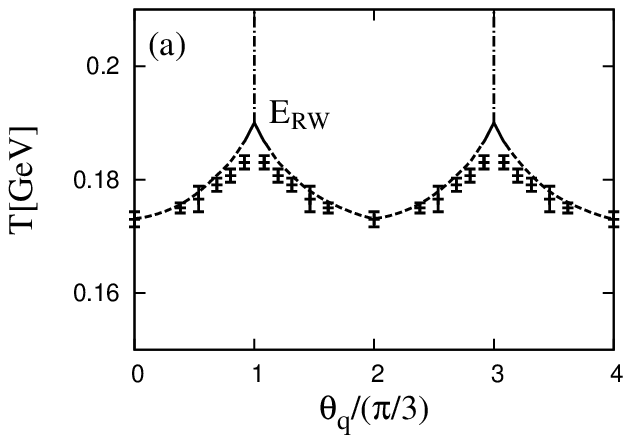}
 \includegraphics[width=0.4\textwidth]{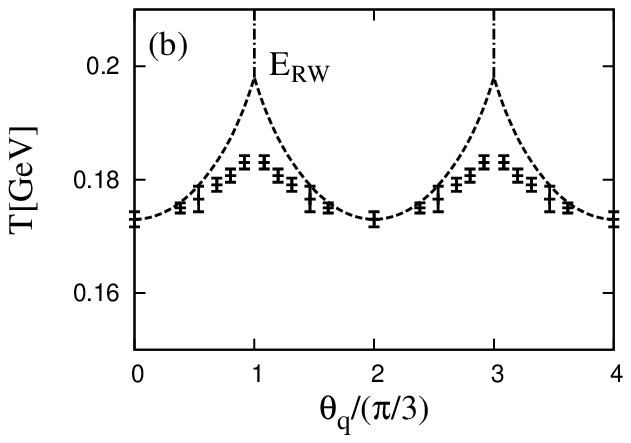}
\end{center}
\caption{
Phase diagrams of the deconfinement and the RW phase transition 
in the $\theta_{\rm q}-T$ plane 
with RRW-type $U_{\Phi}$~\cite{Rossner} 
(panel (a)) and F-type $U_{\Phi}$~\cite{Fukushima} (panel (b)). 
The solid (dashed) line denotes the first-order (crossover) deconfinement 
transition, and the dot-dashed line does the RW transition. 
Point ${\rm E_{RW}}$ is the endpoint of the RW transition. 
Lattice data (+) are taken from~\cite{LQCD-1}.}
\label{Fig2}
\end{figure}
%%%%%%%%%%%%%%%%%%%%%%%%%%%%%%%%%%%%%%%%%%%%%%%%%%%%%%%%%%%%

The phase diagrams of the deconfinement and the RW phase transition in the 
$\theta_{\rm q}-T$ plane by using the PNJL models with 
RRW-type $U_{\Phi}$~\cite{Rossner} and F-type $U_{\Phi}$~\cite{Fukushima} are 
shown in Fig.~\ref{Fig2} (a) and (b), respectively. 
Thus, the PNJL model with  RRW-type $U_{\Phi}$ reproduces 
LQCD data~\cite{LQCD-1} at finite $\theta_{\rm q}$, but the model with F-type 
$U_{\Phi}$ doesn't. 
In this sense, the PNJL model with RRW-type $U_{\Phi}$ calculation is more 
reliable.

%%%%%%%%%%%%%%%%%%%%%%%%%%%%%%%%%%%%%%%%%%%%%%%%%%%%%%%%%%%%
\begin{figure}[htbp]
\begin{center}
 \includegraphics[width=0.40\textwidth]{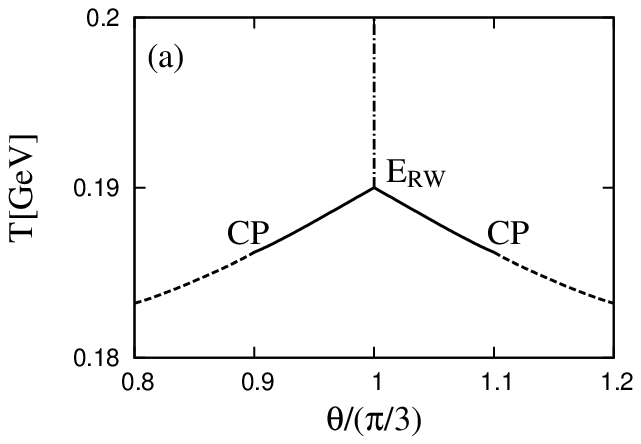}
 \includegraphics[width=0.40\textwidth]{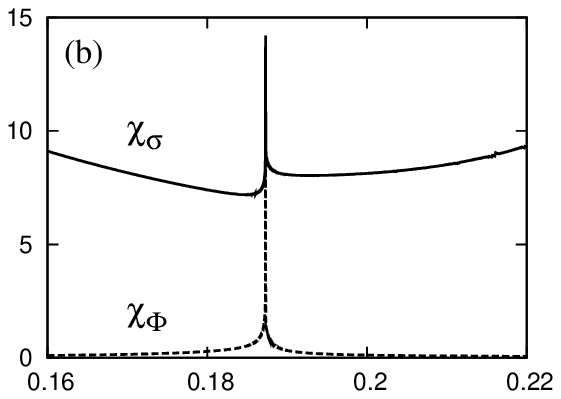}
\end{center}
\caption{
(a) The phase structure near ${\rm E_{RW}}$ with RRW-type $U_{\Phi}$ 
is magnified. 
The solid (dashed) line denotes the first-order (crossover) 
deconfinement transition, and the dot-dashed line does the RW transition. 
Points ${\rm E_{RW}}$ and CP are an endpoint of the RW transition and critical 
endpoints, respectively. 
(b) $T$ dependence of 
the chiral and Polyakov-loop susceptibilities, $\chi_{\sigma}$ and 
$\chi_{\Phi}$, at the point CP. 
}
\label{Fig3}
\end{figure}
%%%%%%%%%%%%%%%%%%%%%%%%%%%%%%%%%%%%%%%%%%%%%%%%%%%%%%%%%%%%

The phase diagram for RRW-type $U_{\Phi}$ near ${\rm E_{RW}}$ is 
magnified in Fig.~\ref{Fig3} (a). 
The RW endpoint is first order for RRW-type $U_{\Phi}$, but it's second 
order for F-type $U_{\Phi}$~\cite{Sakai-2}. 
Thus, the order of the deconfinement phase transition near the RW endpoint 
strongly depends on $U_{\Phi}$ taken. 
The result of the PNJL calculation with RRW-type $U_{\Phi}$ is consistent with 
the LQCD data~\cite{LQCD-2} 
where the order of the RW phase transition at ${\rm E_{RW}}$ is first order 
for small quark mass. 
Point ${\rm E_{RW}}$ is the triple point where the three 
first-order lines meet. Thus, there are two critical endpoints (CP) for each 
triple point ; CP is a point where the crossover and the first order lines 
meet. 
Figure~\ref{Fig3} (b) shows the chiral and the Polyakov loop susceptibilities, 
$\chi_{\sigma}$ and $\chi_{\Phi}$, as a function of $T$ near CP. 
The susceptibilities are divergent at CP. 
Hence, the chiral and deconfinement transitions are second order at CP.

\section{Imaginary Isospin Chemical Potential}

LQCD has no sign problem at both real and imaginary $\mu_{\rm I}$. 
Recently, LQCD data were measured there and also in the case where both 
$\mu_{\rm I}$ and $\mu_{\rm q}$ are imaginary~\cite{LQCD-3}.

In the chiral limit, QCD has the chiral $SU_{\rm L}(2)\times SU_{\rm R}(2)$ 
symmetry when $\mu_{\rm I}=0$. 
However, at $\mu_{\rm I}\ne 0$ this symmetry is reduced to 
$U_{\rm I_3}(1)\times U_{\rm AI_3}(1)$, where $U_{\rm I_3}(1)$ is the isospin 
subgroup and $U_{\rm AI_3}(1)$ is the axial isospin subgroup. 
In the case $m_u=m_d\ne 0$, only the $U_{\rm I_3}(1)$ symmetry survives.
When QCD vacuum keeps the $U_{\rm v}(1)$ and $U_{\rm I_3}(1)$ symmetries, 
the baryon charge $B=V\langle\hat{B}\rangle$ is either zero or integer and the isospin 
charge $I_3=V\langle\hat{I}_3\rangle$ is also 
either zero or half-integer, where 
$\hat{B}=\bar{q}\gamma_4 q, \hat{I}_3=\bar{q}\gamma_4 I_3 q$ and $V$ is the volume. 
In the partition function $Z$,  the baryon- and the isospin-charge operator appear 
through the form $\exp[V(2i\theta_{\rm I}\hat{I}_3 + i \theta_{\rm q}\hat{B})]$
where $\mu_{\rm q, I}=iT\theta_{\rm q, I}$. 
Therefore, $Z$ has the periodicity 
$Z(\theta_{\rm q}, \theta_{\rm I})=Z(\theta_{\rm q}, \theta_{\rm I}+2\pi)$. 
In the isospin symmetric limit $m_u=m_d$, $Z$ is invariant under the interchange 
$u \leftrightarrow d$, i.e. $\theta_{\rm I}\rightarrow-\theta_{\rm I}$.
Hence, $Z$ is invariant under charge conjugation, both 
$\theta_{\rm q}\rightarrow-\theta_{\rm q}$ and 
$\theta_{\rm I}\rightarrow-\theta_{\rm I}$.
Furthermore we have proved that $Z$ has the RW periodicity at 
$\theta_{\rm I}\ne 0$~\cite{Sakai-2}. 
All the relations are summarized as
%%%%%%%%%%%%%%%%%%%%%%%%%%%%%%%%%%%%%%%%%%%%%%%%%%%%%%%%%%%%%%%%%%
\begin{eqnarray}
Z(\theta_{\rm q}, \theta_{\rm I})=Z(\pm\theta_{\rm q}, \mp\theta_{\rm I})=
Z(\theta_{\rm q}, \theta_{\rm I}+2\pi)=Z(\theta_{\rm q}+2\pi/3, \theta_{\rm I}).
\label{Eq3}
\end{eqnarray}
%%%%%%%%%%%%%%%%%%%%%%%%%%%%%%%%%%%%%%%%%%%%%%%%%%%%%%%%%%%%%%%%%%

Meanwhile, if the pion condensation occurs, the $U_{\rm I_3}(1)$ symmetry is 
spontaneously broken and the isospin charge is neither zero nor 
half-integer anymore. In this situation, QCD vacuum doesn't have the 
periodicities (\ref{Eq3}). 
We have proved that the pion condensation doesn't take place at imaginary 
$\mu_{\rm I}$~\cite{Sakai-2}. 
This can be understood intuitively. 
For real $\mu_{\rm I}$, the Bose-Einstein distribution function has an 
infrared divergence at $\mu_{\rm I}\ge m_{\pi}/2$. 
This induces the Bose-Einstein Condensation, that is, the pion condensation. 
For imaginary $\mu_{\rm I}$, such a divergence never happens and hence no pion 
condensation occurs.
As a result of this fact, $Z$ has all the discrete symmetries (\ref{Eq3}).

%%%%%%%%%%%%%%%%%%%%%%%%%%%%%%%%%%%%%%%%%%%%%%%%%%%%%%%%%%%%
\begin{figure}[htbp]
\begin{center}
 \includegraphics[width=0.32\textwidth]{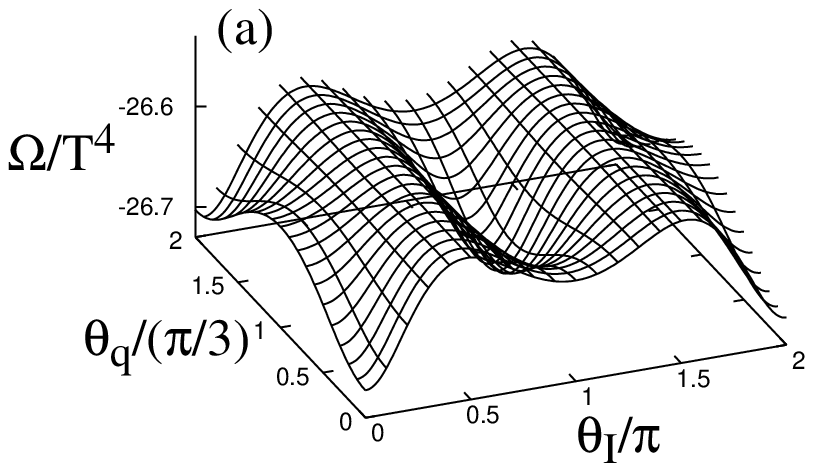}
 \includegraphics[width=0.32\textwidth]{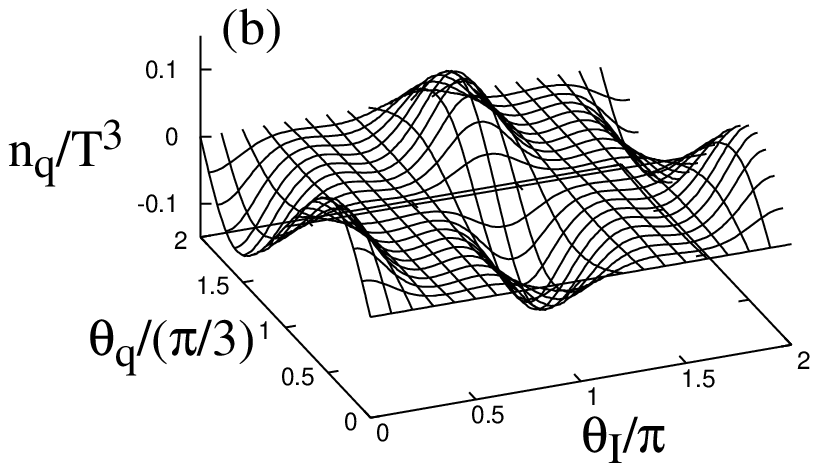}
 \includegraphics[width=0.32\textwidth]{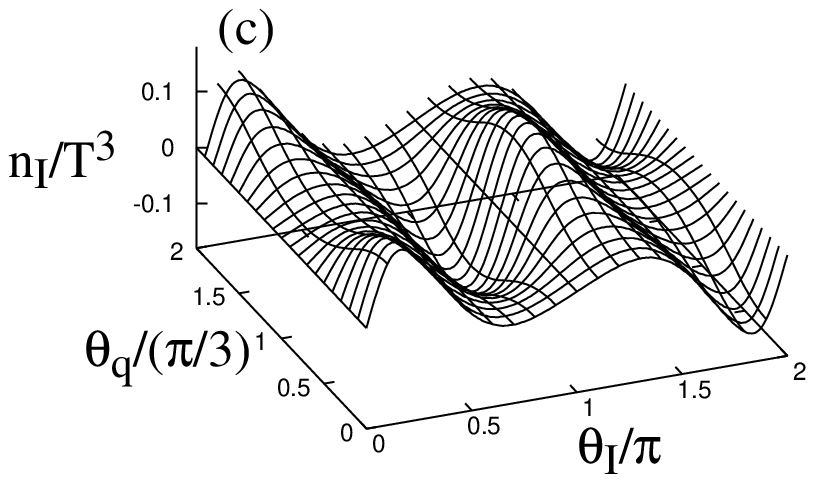}
 \includegraphics[width=0.32\textwidth]{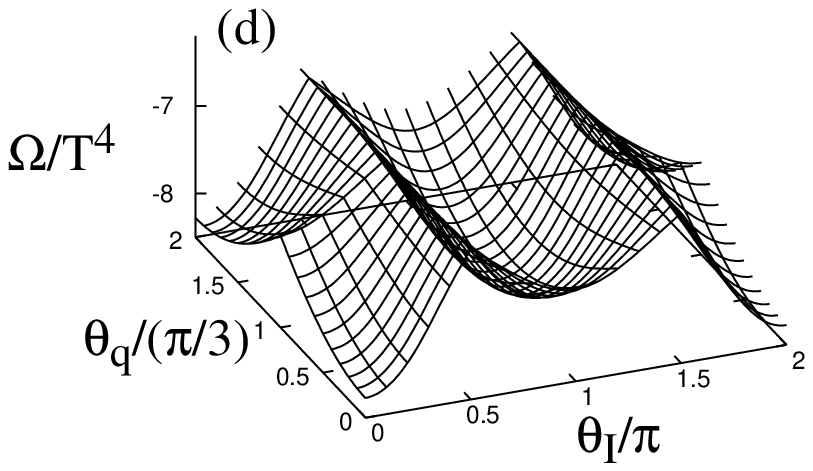}
 \includegraphics[width=0.32\textwidth]{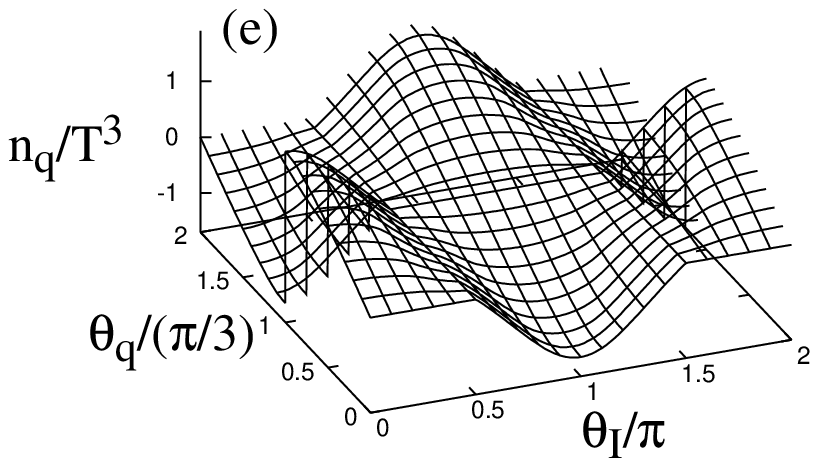}
 \includegraphics[width=0.32\textwidth]{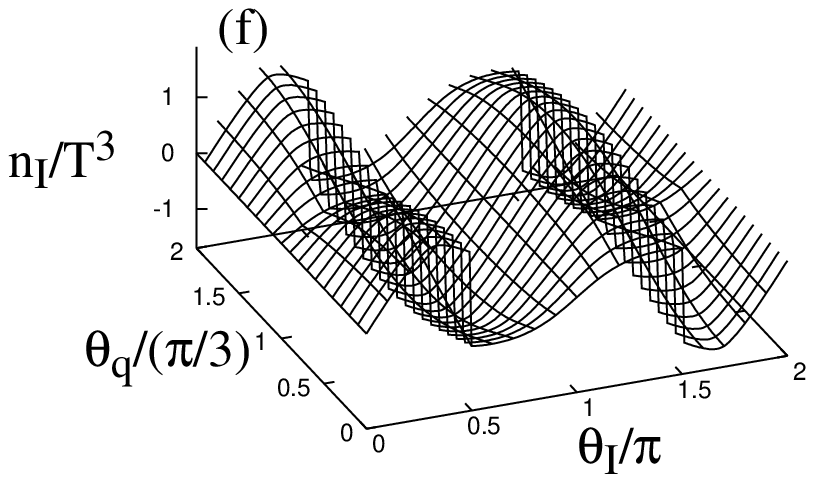}
\end{center}
\caption{
$\Omega/T^4$, $n_{\rm q}/T^3$ and $n_{\rm I}/T^3$ 
as a function of $\theta_{\rm q}$ and $\theta_{\rm I}$. 
Panels (a), (b) and (c) correspond to $T=175$~MeV, 
while panels (d), (e) and (f) to $T=250$~MeV.}
\label{Fig4}
\end{figure}
%%%%%%%%%%%%%%%%%%%%%%%%%%%%%%%%%%%%%%%%%%%%%%%%%%%%%%%%%%%%

The absence of the pion condensation at imaginary $\mu_{\rm I}$ is true in 
the PNJL model~\cite{Sakai-2}. 
The PNJL thermodynamic potential at $\mu_{\rm I}\ne 0$ in the MFA is 
%%%%%%%%%%%%%%%%%%%%%%%%%%%%%%%%%%%%%%%%%%%%%%%%%%%%%%%%%%%%%%%%%%
\begin{eqnarray}
\Omega=-2\int\frac{d^3{\bf p}}{(2\pi)^3}\sum_{f=\pm 1}
\Bigl[3\epsilon_{f}({\bf p}) + T\sum_{\lambda=\pm 1}
{\rm ln~det_c}
(1+L^{\lambda}e^{-\epsilon_{f}({\bf p})/T+i\lambda\theta_{\rm q}})
\Bigr]+G_{\rm s}(\sigma^2+\pi^2)+U_{\Phi},~~~
\end{eqnarray}
%%%%%%%%%%%%%%%%%%%%%%%%%%%%%%%%%%%%%%%%%%%%%%%%%%%%%%%%%%%%%%%%%%
where $\epsilon_\pm=\sqrt{(\epsilon({\bf p})\pm\mu_{\rm I})^2+N^2}$, 
$N=2G_{\rm s}\pi$. 
The pion condensate $\pi=\langle\bar{q}i\gamma_5\tau_1q\rangle$ is an order 
parameter of the spontaneous breakings of the $U_{\rm I_3}(1)$ symmetry. 
When there is no pion condensation, $\Omega$ is reduced to a simpler form
%%%%%%%%%%%%%%%%%%%%%%%%%%%%%%%%%%%%%%%%%%%%%%%%%%%%%%%%%%%%%%%%%%
\begin{eqnarray}
\Omega=-2\int\frac{d^3{\bf p}}{(2\pi)^3}
\Bigl[6\epsilon({\bf p}) + T\sum_{\lambda,~f=\pm 1}
{\rm ln~det_c}(1+L^{\lambda}e^{-\epsilon({\bf p})/T+i\lambda\theta_{\rm q}
+if\theta_{\rm I}})\Bigr]+G_{\rm s}\sigma^2+U_{\Phi}, 
\end{eqnarray}
%%%%%%%%%%%%%%%%%%%%%%%%%%%%%%%%%%%%%%%%%%%%%%%%%%%%%%%%%%%%%%%%%%
which is invariant under the extended 
${\mathbb Z}_3$ transformation (\ref{Eq2-3}), therefore $\Omega$ has the RW 
periodicity. 
The potential $\Omega$ has also the periodicity of 
$\theta_{\rm I}\rightarrow\theta_{\rm I}+2\pi$. 
Furthermore $\Omega$ is invariant under the transformation, 
$\theta_{\rm I}\rightarrow-\theta_{\rm I}$, and 
also under the transformation, $\theta_{\rm q}\rightarrow-\theta_{\rm q}$ and 
$L^{\pm}\rightarrow L^{\mp}$. 
These properties guarantee that 
the PNJL model possesses all the symmetries in (\ref{Eq3}), and the model 
reproduces LQCD data~\cite{LQCD-3} qualitatively at imaginary $\mu_{\rm I}$ 
and $\mu_{\rm q}$.

Figure~\ref{Fig4} shows $\Omega/T^4$, ${\rm Im}[n_{\rm q}]/T^3$ and 
${\rm Im}[n_{\rm I}]/T^3$ as a function of $\theta_{\rm q}$ and 
$\theta_{\rm I}$ in the cases of $T=175$ and $250$~MeV. 
Symmetries (\ref{Eq3}) are seen in Fig.~\ref{Fig4}. 
This result is consistent with LQCD ones~\cite{LQCD-3}. 
If the pion condensate is nonzero, symmetries (\ref{Eq3}) break down. 
Hence, the fact that LQCD has symmetries (\ref{Eq3}) means 
that the pion condensation doesn't occur also in LQCD. 
As shown in Fig.~\ref{Fig2} (a) for $\theta_{\rm I}=0$, at temperature above 
$T_{\rm RW}=190$~MeV, there is the RW phase transition at 
$\theta_{\rm q}=\pi/3$ mod $2\pi/3$, where 
$n_{\rm q}=-{\rm d}\Omega/{\rm d}(iT\theta_{\rm q})$ is discontinuous. 
In Fig.~\ref{Fig4}, $T=175$ and $250$~MeV are typical temperatures below 
and above $T_{\rm RW}$, respectively. 
For any temperature, the RW periodicity is seen. 
Below $T_{\rm RW}$, these quantities are smooth at any $\theta_{\rm q}$ and 
$\theta_{\rm I}$. 
In contrast, above $T_{\rm RW}$, 
$\Omega$ and $n_{\rm I}$ have cusps at $\theta_{\rm q}=\pi/3$ mod 
$2\pi/3$, while $n_{\rm q}$ is discontinuous there. 
The discontinuity means the RW phase transition.
Eventually, the transition occurs at $\theta_{\rm q}=\pi/3$ mod $2\pi/3$ 
when $0\le\theta_{\rm I}<\pi/2$ and $\pi<\theta_{\rm I}\le 2\pi$, 
and at $\theta_{\rm q}=0$ mod $2\pi/3$ when $\pi/2\le\theta_{\rm I}\le3\pi/2$~\cite{Sakai-2}.

%%%%%%%%%%%%%%%%%%%%%%%%%%%%%%%%%%%%%%%%%%%%%%%%%%%%%%%%%%%%
\begin{figure}[htbp]
\begin{center}
 \includegraphics[width=0.23\textwidth,angle=-90]{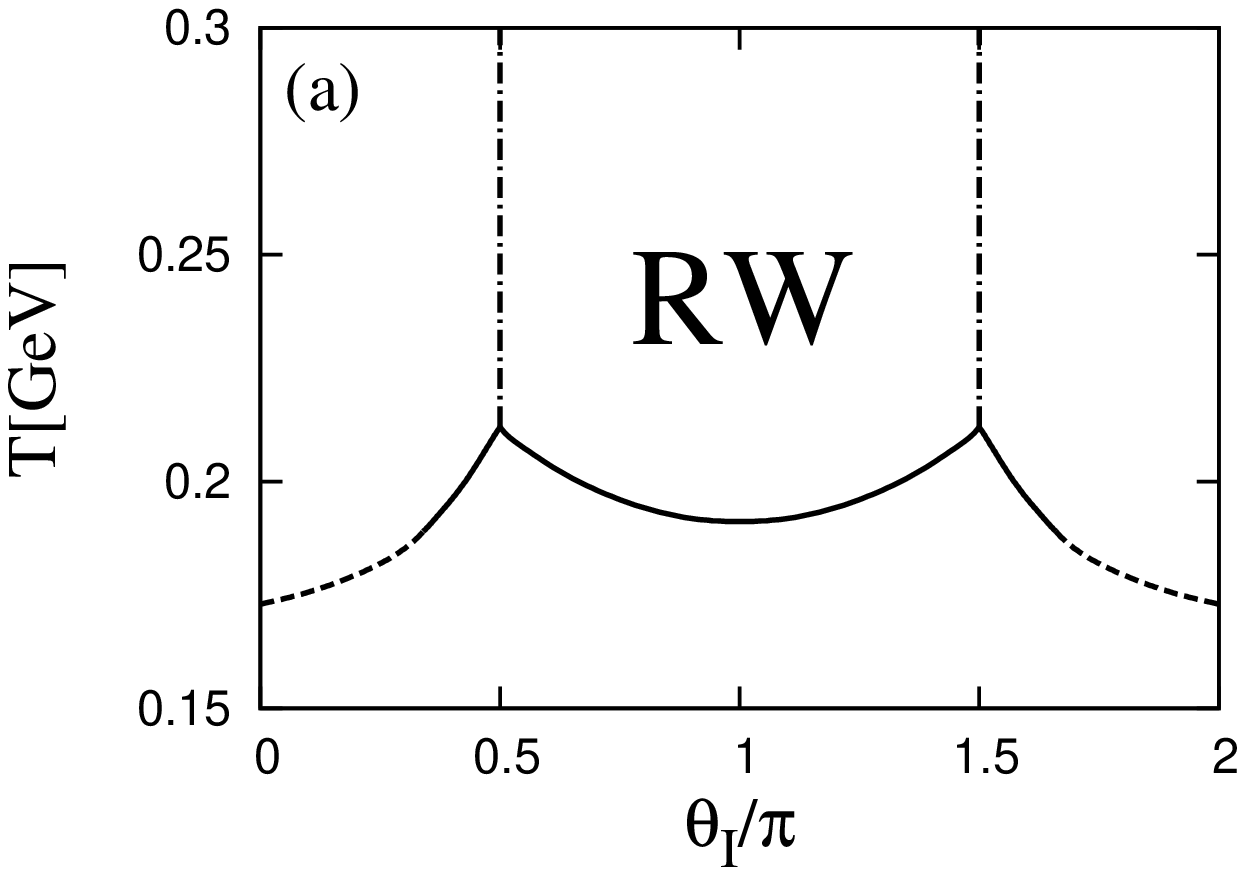}
 \includegraphics[width=0.23\textwidth,angle=-90]{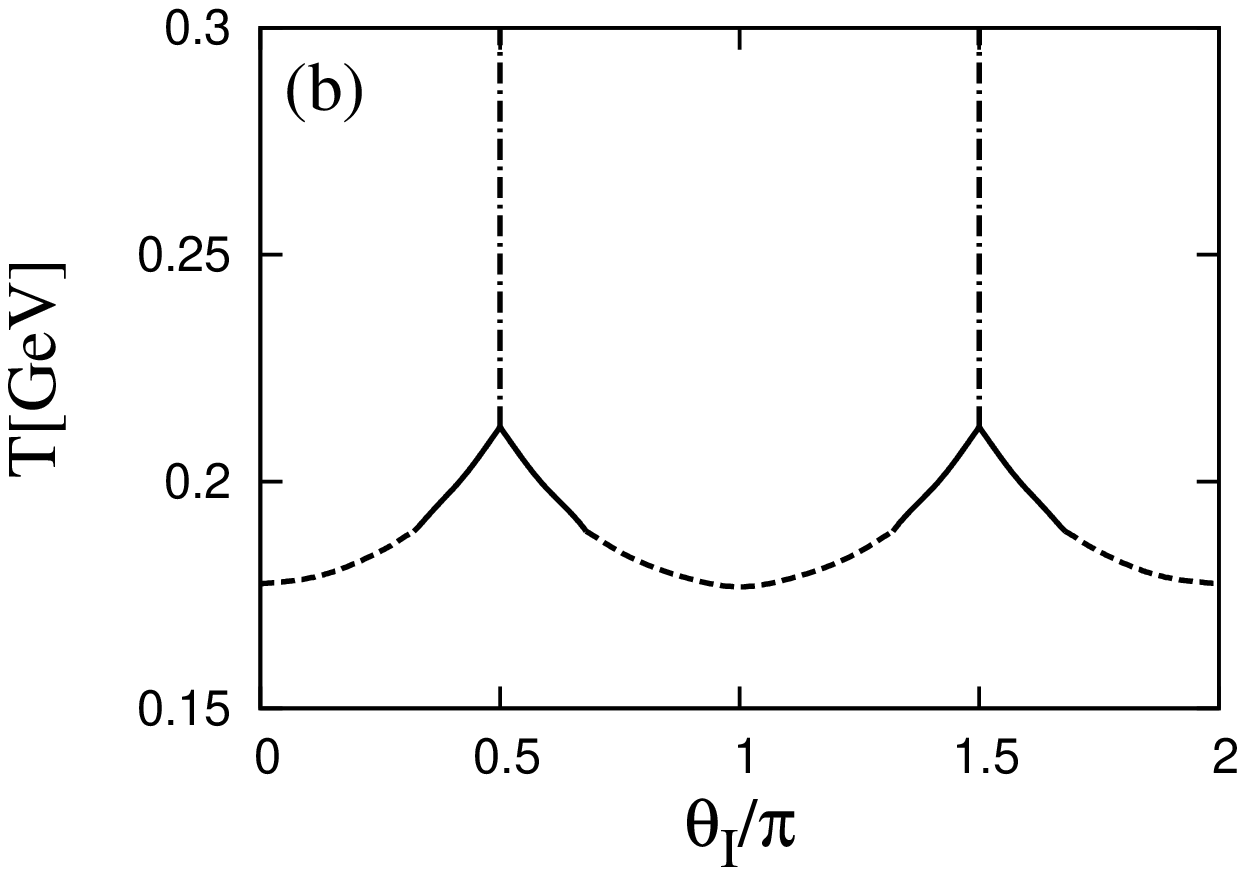}
 \includegraphics[width=0.23\textwidth,angle=-90]{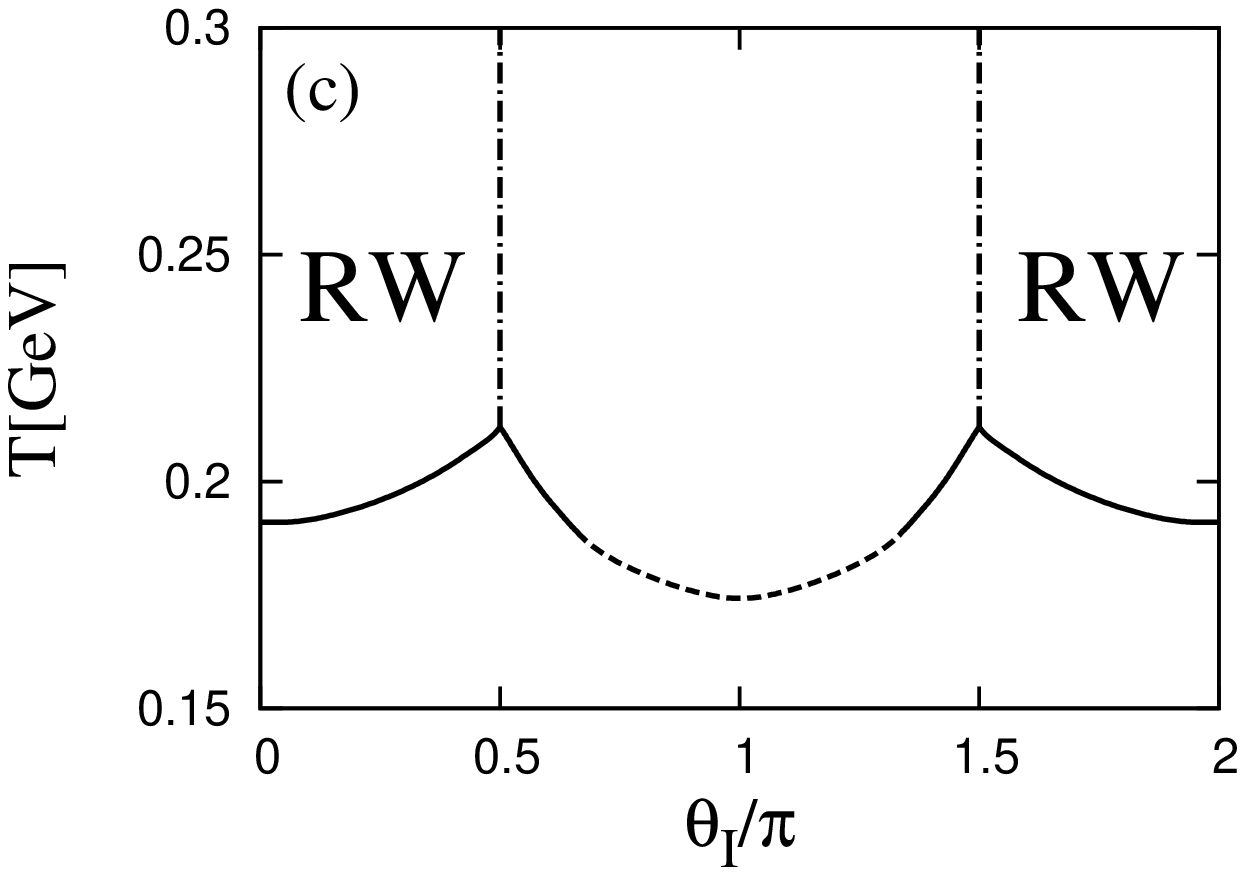}
\end{center}
\caption{
Phase diagram of the deconfinement phase transition in the $\theta_{\rm I}-T$ 
plane. Panels (a), (b) and (c) correspond to $\theta_{\rm q}=0, \pi/6$ and 
$\pi/3$, respectively. 
The solid (dashed) line denotes the first-order (crossover) transition. 
The area labeled by 'RW' between the two dot-dashed 
lines represents the region where the RW phase transition occurs.}
\label{Fig5}
\end{figure}
%%%%%%%%%%%%%%%%%%%%%%%%%%%%%%%%%%%%%%%%%%%%%%%%%%%%%%%%%%%%

Figure~\ref{Fig5} shows the phase diagram of the deconfinement phase 
transition in the $\theta_{\rm I}-T$ plane. 
Near $\theta_{\rm I}=\pi/2$ mod $\pi$, the deconfinement phase transition is 
first order in all panel (a)-(c). Near $\theta_{\rm I}=\pi$ mod $\pi$, 
the deconfinement phase transition is first order at $\theta_{\rm q}=0$, 
but crossover at $\theta_{\rm q}=\pi/6$ and $\pi/3$. 
The RW phase transition occurs in the area labeled by 'RW' between the two 
dot-dashed lines.

Quantitative comparison of the PNJL model with LQCD data~\cite{LQCD-3} is 
made at $T\le T_{\rm c}$ by using the hadron resonance gas (HRG) model that 
can reproduce the LQCD data there. 
We have shown~\cite{Sakai-2} that the PNJL model reproduces the LQCD data 
for the oscillatory patterns. 
For the magnitudes, meanwhile, the PNJL model underestimates the LQCD result. 
This discrepancy is understandable as follows. 
Below $T_{\rm c}$, hadronic excitations are important, but such 
an effect is not included in the MFA. 
By adding the hadronic correction to the PNJL model, the model agrees with the 
LQCD~\cite{Sakai-2}. 
The HRG model works well at $T<T_{\rm c}$, but not 
at $T>T_{\rm c}$; especially the HRG model doesn't reproduce the RW phase 
transition. In contrast, the PNJL model with the hadronic correction works 
both below and above $T_{\rm c}$.

\section{Real Isospin Chemical Potential}

\begin{figure}[htbp]
\begin{center}
\begin{minipage}{0.47\textwidth}
 \includegraphics[width=0.90\textwidth]{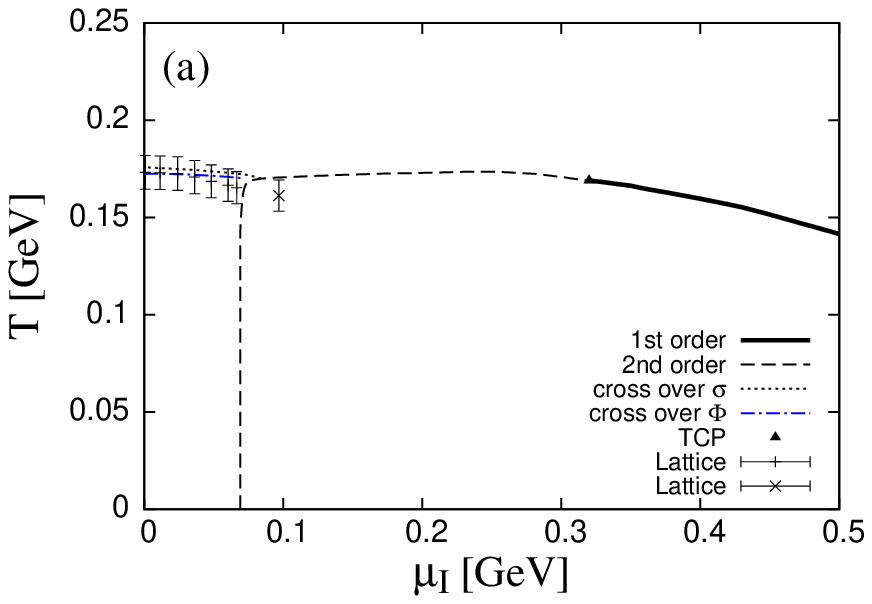}
\end{minipage}
\begin{minipage}{0.47\textwidth}
\hspace{-0.5cm}
 \includegraphics[width=1.20\textwidth]{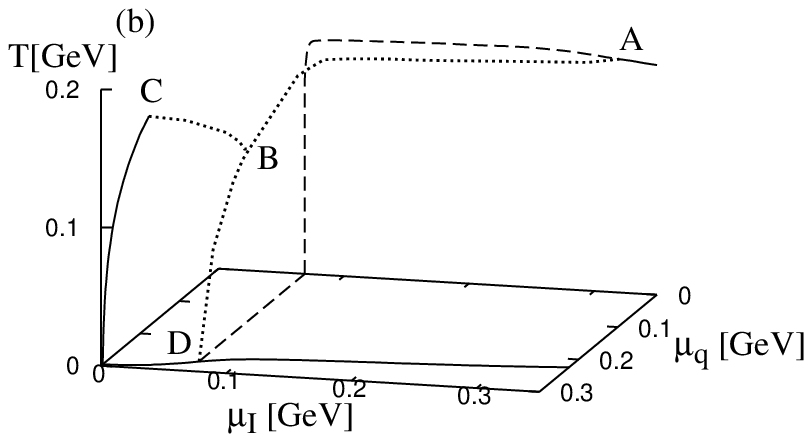}
\end{minipage}
\end{center}
\vspace{-1cm}
\caption{
(a) Phase diagram in the $\theta_{\rm I}-T$ plane at $\theta_{\rm q}=0$ 
with the eight-quark interaction. 
The thick-solid (dashed) line denotes a first-order 
(second-order) pion-superfluidity phase transition. 
The dot-dashed (dotted) line denotes a deconfinement (chiral) crossover 
transition. Lattice data are taken from~\cite{LQCD-4}. 
(b) Phase diagram in the $\mu_{\rm I}-\mu_{\rm q}-T$ space 
with the eight-quark interaction. Line ABC denotes the chiral CEP, ABD line 
does the pion-superfluid TCP. 
The CEP and the TCP coexist on line AB. 
The solid (dashed) line denotes the first (second) order transition. 
}
\label{Fig6}
\end{figure}

LQCD data are available at real $\mu_{\rm I}$ and $\mu_{\rm q}=0$
~\cite{LQCD-4}. 
The scalar-type eight-quark interaction is necessary to reproduce LQCD data 
at imaginary $\mu_{\rm q}$~\cite{Sakai-1}. 
Figure~\ref{Fig4} (a) shows the phase diagram of the PNJL model with the 
scalar-type eight-quark interaction in the $\mu_{\rm I}-T$ plane at 
$\mu_{\rm q}=0$.
The PNJL model with the eight-quark interaction is also consistent with the 
LQCD at $\mu_{\rm I}\ne 0$~\cite{Sasaki}. 
There is a tricritical point (TCP) where the first-order pion-superfluidity 
phase transition line is connected to the second-order phase transition. 
The critical points such as CEP and TCP are important as indicators
of the chiral and pion-superfluidity phase transitions at compact stars and 
laboratory experiments where $\mu_{\rm I}$ is nonzero generally. 
The TCP in the $\mu_{\rm I}-T$ plane at $\mu_{\rm q}=0$ is connected to the 
CEP in the $\mu_{\rm q}-T$ plane at $\mu_{\rm I}=0$ in the 
$\mu_{\rm q}-\mu_{\rm I}-T$ space~\cite{Sasaki}, as shown 
in Fig.~\ref{Fig6} (b). 

%%%%%%%%%%%%%%%%%%%%%%%%%%%%%%%%%%%%%%%%%%%%%%%%%%%%%%%%%%%%
%%%%%%%%%%%%%%%%%%%%%%%%%%%%%%%%%%%%%%%%%%%%%%%%%%%%%%%%%%%%

%%%%%%%%%%%%%%%%%%%%%%%%%%%%%%%%%%%%%%%%%%%%%%%%%%%%%%%%%%%%%%%%%%%%%%%%%%%%%%%%%%%%% References 
%%%%%%%%%%%%%%%%%%%%%%%%%%%%%%%%%%%%%%%%%%%%%%%%%%%%%%%%%%%%%%%%%%%%%%%%%%%%%%%%

\end{document}